\documentclass[useAMS,usenatbib,usegraphicx,twocolumn]{mn2e}
\usepackage[figuresright]{rotating}
\usepackage{amsmath,amssymb}

\title[Orbital structure in barred spiral galaxies]{Orbital structure in $N$-Body models of barred-spiral galaxies}
\author[M. Harsoula, and C. Kalapotharakos]
       {M. ~Harsoula and C. ~Kalapotharakos\\
        Research Center for Astronomy,
           Academy of Athens, Soranou Efesiou 4, GR-115 27 Athens, Greece\\
           e-mail: mharsoul@academyofathens.gr,
           ckalapot@phys.uoa.gr}
\date{Released 2007 April 12}
\pagerange{\pageref{firstpage}--\pageref{lastpage}} \pubyear{2007}

\def\LaTeX{L\kern-.36em\raise.3ex\hbox{a}\kern-.15em
    T\kern-.1667em\lower.7ex\hbox{E}\kern-.125emX}

\begin{document}

\maketitle

\label{firstpage}

\begin{abstract}
We study the orbital structure in a series of self-consistent
$N$-body configurations simulating rotating barred galaxies with
spiral and ring structures. We perform frequency analysis in order
to measure the angular and the radial frequencies of the orbits at
two different time snapshots during the evolution of each $N$-body
system. The analysis is done separately for the regular and the
chaotic orbits. We thereby identify the various types of orbits,
determine the shape and percentages of the orbits supporting the bar
and the ring/spiral structures, and study how the latter quantities
change during the secular evolution of each system. Although the
frequency maps of the chaotic orbits are scattered, we can still
identify concentrations around resonances. We give the distributions
of frequencies of the most important populations of orbits. We
explore the phase space structure of each system using projections
of the 4D surfaces of section. These are obtained via the numerical
integration of the orbits of test particles, but also of the real
$N$-body particles. We thus identify which domains of the phase
space are preferred and which are avoided by the real particles. The
chaotic orbits are found to play a major role in supporting the
shape of the outer envelope of the bar as well as the rings and the
spiral arms formed outside corotation.

\end{abstract}


\section{Introduction}

The study of the orbits in barred galaxies begun in the late 70's
\citep{b45,b46} when the properties of the orbits in simple models
of barred galaxies were derived theoretically by use of a `third
integral' of motion besides the Hamiltonian. \citet{b47} made an
extended study of the {\it periodic orbits} in weak and strong
bars, establishing the standard nomenclature thenceforward (see
\citet{b103} and \citet{b43} for a review of the various families
of periodic orbits both inside and outside corotation). Many works
have been devoted to the study of both periodic and non-periodic
orbits in 2D \citep[e.g.][]{b56,b18,b104,b27,b44} or 3D models
\citep[e.g.][]{b57,b50,b22,b23,b24,b25,b21}. In most studies an
`ad hoc' choice of model is made for the gravitational potential.
However, some studies have addressed the question of the orbital
structure in $N$-body systems of barred galaxies
\citep[e.g.][]{b32,b33,b34,b35,b36,b16} in which, by definition,
the orbits support the system self-consistently. Studies of the
latter type are, in fact, necessary in order to identify which
families of orbits are most relevant to the maintenance of
self-consistency, i.e. to the production, by the orbits, of
`response' patterns matching those of the imposed
potential/density field.

\begin{figure}
\centering
\includegraphics[width=8.5cm]
{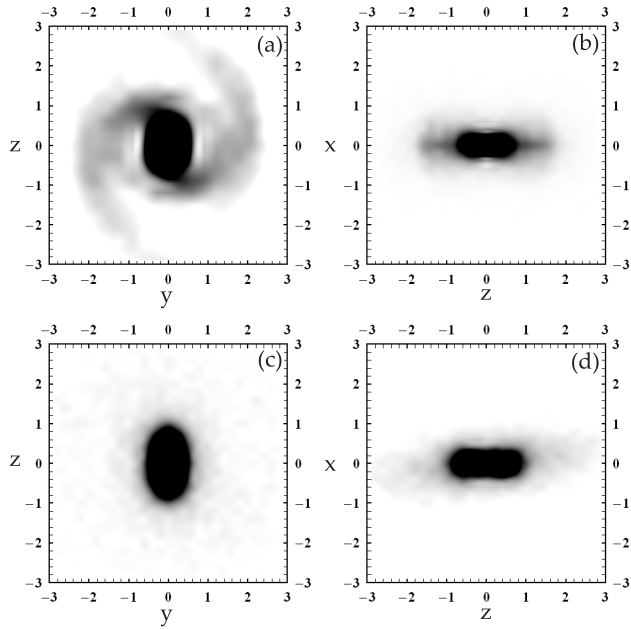} \caption{The face-on and edge-on density profiles (left
and right column respectively) for the experiment QR3 at
$t=20T_{hmct}$ and at $t=300T_{hmct}$ (first and second row
respectively).} \label{fig:1}
\end{figure}

\begin{figure}
\centering
\includegraphics[width=8.5cm]
{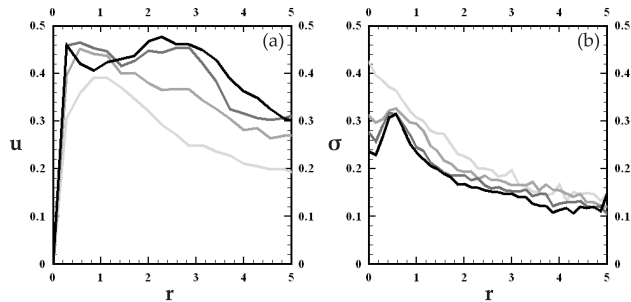} \caption{The rotation curves (a) and the velocity
dispersions (b), for all the experiments at $t=20T_{hmct}$, along
the major axis considering the line-of-sight along the middle axis.
Lines from light gray to black correspond to the experiments from
QR1 to QR4 respectively.} \label{fig:2}
\end{figure}

By constructing self-consistent models simulating real
barred-spiral galaxies, \citet{b49} pointed out the role of the
{\it chaotic orbits} in supporting the inner parts of the spiral
arms emanating beyond the bar. The orbits considered by Kaufmann
\& Contopoulos belong to the so-called `hot population' (Sparke
and Sellwood 1987), i.e. they wander stochastically, partly inside
and partly outside corotation. On the other hand, the same authors
found that the outer parts of the spiral arms are supported mainly
by regular orbits. The study of the role of the chaotic orbits in
the spiral structure was pursued by \citet{b31}, using self
consistent $N$-body simulations of barred galaxies. These authors
found long living spiral arms composed almost entirely by chaotic
orbits. A new mechanism was proposed
\citep{b10,b100,b39,b101,b140} according to which the invariant
manifolds of the short period unstable periodic orbits around the
stationary Lagrangian points $L_1$ and $L_2$ are responsible for
the long-term support of the spiral structure. On the other hand
\citet{b40} argued that the innermost parts of the spiral arms,
formed as a continuation of the bar, a little inside corotation,
are due to chaotic orbits exhibiting, for long time intervals, a
4:1 resonance orbital behavior.

In \citet{b1} (hereafter paper I) a detailed investigation was made
of the orbital structure in one system belonging to the series of
$N$-body experiments reported in \citet{b31}. This particular
simulation represents a barred galaxy with nearly no spiral
structure. Using i) the method of frequency analysis (Laskar 1990),
and ii) 2D projections of the 4D surface of sections, obtained via
the numerical integration both for test particles and of the real
$N$-body particles, the relative percentages were found of a number
of distinct populations which support the bar. The analysis was
carried out separately for the regular and the chaotic orbits. A
significant fraction of chaotic orbits were found to lie inside
corotation, thus supporting the shape of the bar. Moreover it was
shown that in a 2D approximation in which the system's thickness is
ignored, the chaotic orbits are limited by tori or cantori so that
many orbits remain confined inside limited domains outside
corotation for times comparable to the Hubble time. Arnold diffusion
through the third dimension was found ineffective over such time
periods, thus the results are applicable in the 3D case as well.

In the present paper we extend the above study in the whole series
of $N$-body experiments reported in \citet{b31}, focusing in
particular in those systems which simulate secularly evolving barred
galaxies with significant spiral structure. The simulations
represent self-consistent systems of a wide range of different
pattern speeds of the bar. One main goal is to determine the
dependence of the percentage of the different populations (particles
in different types of orbits) on the value of the pattern speed. It
should be stressed that the pattern speed of one system slows down
as the system evolves secularly. To account for this effect, we
study the orbital structure in each experiment at two distinct and
well separated time snapshots, one close to the beginning of the
simulation and one close to the final state (corresponding to one
Hubble time evolution). This analysis yields the changes induced
upon the different types of orbits, as well as upon the
distributions of the $N$-body particles in the orbital space, by the
secular evolution of each galaxy towards an equilibrium state.

In order to accomplish the orbital classification, we implement
the method of frequency analysis both for the regular and for the
chaotic orbits of each system. The method of frequency analysis
was first introduced in the study of bar like potentials
\citep{b150} and in the study of resonances between planetary
orbits in the Solar system \citep{b2}, and within the framework of
galactic dynamics it was implemented in the orbital study of
elliptical galaxies \citep{b60,b59,b3,b58,b4,b5}, or barred
galaxies \citep{b14,b17,b1,b16}. Here we extend the implementation
of the method to the problem of spiral structure as well.

\begin{figure*}
\centering
\includegraphics[width=12.0cm]
{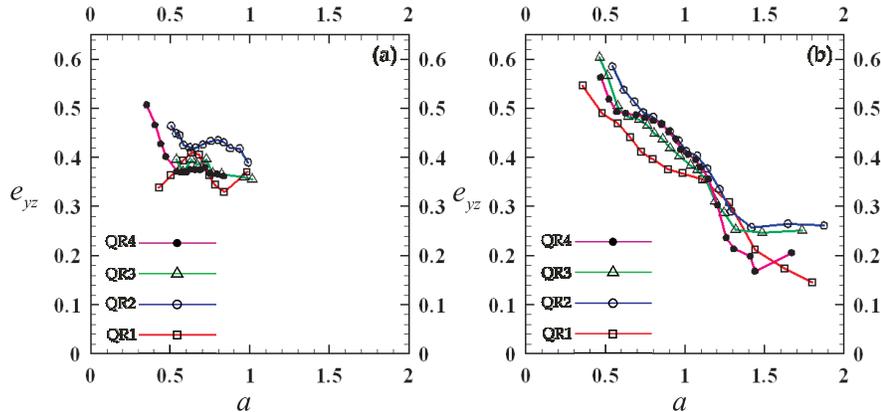} \caption{The ellipticity profile of the bar's isophotes
for the four experiments (a) $t=20T_{hmct}$, (b) $t=300T_{hmct}$.}
\label{fig:3}
\end{figure*}

Besides identifying various types of regular orbits, concentrated
around well known resonances, we find that many chaotic orbits,
especially those being confined inside corotation, also exhibit a
large degree of concentration around specific resonances. The main
difference is that the chaotic orbits yield more diffused
frequency maps than the regular orbits. Most chaotic orbits inside
corotation can be further characterized as `weakly chaotic', i.e.
they have low values of the Lyapunov exponents. These orbits
support the outer region of the bar. On the other hand, the
chaotic orbits outside corotation are in principle capable of
escaping. In most cases, however, the escape time is much longer
than one Hubble time. Thus, in practice many chaotic orbits
outside corotation are also trapped by various resonances
exhibiting `stickiness' along the invariant manifolds of the
unstable periodic orbits near the Lagrangian points $L_1, L_2,
L_4$ and $L_5$. We demonstrate that the different populations of
such orbits are responsible for practically all observed
significant morphological features beyond the bar, and in
particular for the observed rings and/or spiral arms.

\begin{figure}
\centering
\includegraphics[width=6cm]
{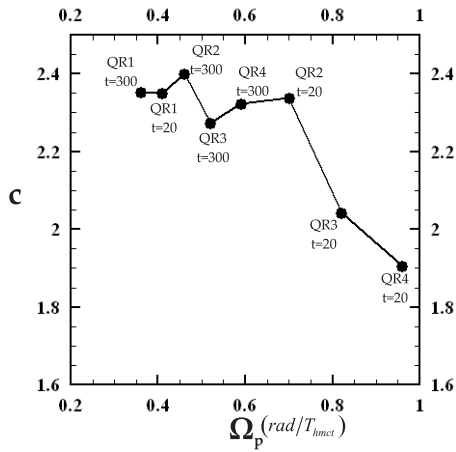} \caption{The $c$-parameter as a function of the pattern
speed of the bar for all the experiments and time snapshots.}
\label{fig:4}
\end{figure}

Further examination of the phase space is performed using
projections of the 4D surface of section of test particles. Using
the same technique for the real $N$-body particles, an important
information is recovered, namely the location of those domains of
the phase space which are preferred and those which are avoided by
the real particles.

The paper is organized as follows. In section 2 we describe briefly
the $N$-body models and discuss the correlation between the pattern
speed and the boxiness of the bar. In section 3.1 we present briefly
the methods used for the classification of the orbits into regular and
chaotic, we introduce the frequency analysis and we give the main
results regarding the detailed orbital structure for all the models.
In section 3.2 we examine the phase space structure through surfaces
of sections using test particles as well as real $N$-body particles.
Finally section 4 summarizes the conclusions of the present study.

\section[]{Description of the models}

The series of simulations referred to in the sequel are described in
detail in \citet{b31}. The experiments were evolved using the
\textbf{S}mooth \textbf{F}ield \textbf{C}ode (hereafter SFC) of
\citet{b37}. The initial conditions were created by introducing
rotation in the so called `Q-model' of \citet{b30} in the way
described below.

A measure of the angular momentum $J$, for a gravitational system
of total mass M and total binding energy E, can be given in terms
of Peebles' (1969) `spin parameter' $\lambda$:
\begin{equation}
\lambda=\frac{J|E|^{1/2}}{GM^{5/2}}
\end{equation}

The Q-model of \citet{b30} has initially an almost zero spin
parameter, simulating an E7 elliptical galaxy \cite[in contrast,
the value of $\lambda$ for, say, the disc of our Galaxy is
$\lambda_G=0.22$, see][]{b102}. In fact, a value of $\lambda$
close to the one observed in real barred-spiral galaxies cannot be
obtained via a purely collisionless evolution of either a
monolithically collapsing object, embedded in the tidal field of
surrounding density perturbations, or a system formed by mergers
between subclumps in a hierarchical clustering process. In order
to produce systems with a $\lambda$ parameter appropriate for
barred galaxies, while running a collisionless simulation,
\citet{b31} introduced a process called `re-orientation of the
velocities'. Starting from an equilibrium triaxial system (e.g.
the Q-system), all the particles' velocity vector components on
the plane defined by the intermediate and long axes of the
triaxial equilibrium figure are reoriented as to render each
particle's velocity vector perpendicular to its position vector.
Furthermore, the re-orientation is such that the new velocity
vector defines a clockwise sense of the particle's orbital
revolution. This process yields the maximum possible rotation of a
system, on the above referenced plane, with respect to the
progenitor system which has the same density and equal
distribution of the velocity moduli. The experiment obtained from
the Q-system via `velocity re-orientation' was called QR1
\citep[see][]{b31}. This was left to run self-consistently, using
an improved SFC version, for one Hubble time. In a similar way,
the experiment QR2 was produced by introducing a new velocity
re-orientation at a snapshot of the QR1 run corresponding to 20
half mass crossing times (hereafter $T_{hmct}$). The experiment
QR3 was produced in the same way from QR2, and so on. We shall
consider four $N$-body experiments of this sequence (QR1 to QR4).
All these experiments have the same number of particles ($\approx
1.5\times10^5$) and the same binding energy but different amounts
of total angular momentum, resulting therefore in different values
of the bar's pattern speed. Further details are provided in
\citet{b31}. It must be pointed out here that in our
self-consistent systems the particles cannot be identified a
priori as belonging to a disc, halo or bulge component. The
systems are flattened as a whole and when viewed edge-on they
present a shape called `thick disc' in \citet{b31}. The dynamical
effects caused by the particles in the `thick disk' were analyzed
in details in \citet{b101}, in subsection 2.1, where it was shown
that the particles with large vertical oscillations create a
gravitational field which partly mimics the effect of a halo up to
the code's truncation radius. Moreover, it is possible to
recognize even the bulge that is created self-consistently in
barred galaxies, using a method proposed by observers, i.e. by
plotting the density profile along and parallel to the major or
minor axis of the bar. The bulge length is marked by the increased
light distribution over the exponential disc, well above the bar
(see Paper I for details).

In each experiment, the time unit is taken equal to the
$T_{hmct}$. A Hubble time corresponds to $\approx300T_{hmct}$. The
length unit is taken equal to the half mass radius (hereafter
$r_{hm}$). Finally, the plane of rotation is the $y-z$ plane
(intermediate-long axes) and the sense of rotation is clockwise.

In Fig.1 the density of the particles of the experiment QR3 is
plotted face-on (Fig.1a,c) and edge-on (Fig.1b,d), for two
different snapshots $t=20T_{hmct}$ (Fig.1a,b) and $t=300T_{hmct}$
(Fig.1c,d). Some other snapshots of these four experiments in the
ordinary space can be seen in fig.1 of \citet{b31}.

Figure 2a shows the rotation curves obtained by calculating the
mean values of the line-of-sight velocity profiles of the
particles when the systems are viewed edge-on and the bar is
side-on (with the line of sight along the middle axis), for all
the experiments from QR1 (light gray line) to QR4 (black line) at
time $t=20T_{hmct}$. In Fig.2b the corresponding velocity
dispersion profiles are plotted.

The slow down of the pattern speed in our experiments (see fig.4a in
Voglis et al. 2006a) is smaller compared with other simulations
using `live halos' (see for example Athanassoula 2003 and
Martinez-Valpuesta et al. 2006). Moreover, the evolution towards an
equilibrium state seems faster. One possible reason is that, in our
simulations, a bar already exists from the beginning of the
calculations $(t=0)$, where rotation has been inserted, via
re-orientation of the velocities' vectors, in all the particles of
the system. Moreover, our systems present an important velocity
dispersion especially in the central region (see Fig.2b) at time
$t=20T_{hmct}$ which is close to the initial conditions, and this
favors the faster slow down of the bar's pattern speed according to
the main conclusion of Athanassoula (2003).

Apart from the QR1 experiment, in which the spiral structure does
not survive for more that $\approx15T_{hmct}$, in all the other
experiments there are $m=2$ spiral modes surviving for 300
$T_{hmct}$, i.e. for one Hubble time. The amplitude of these modes
undergoes oscillations with an overall tendency to decay. However,
during this secular evolution the spiral modes are more prominent at
particular time snapshots. In all three models QR2, QR3 and QR4 the
spiral structures are relatively strong at t=20$T_{hmct}$.

Figure 3 shows the ellipticity $e_{yz}=1-b/a$ as a function of the
major semi-axis $a$ of an isophote ($b$ is the isophote's minor
semi-axis) for the isophotes corresponding to the projected surface
density on the $y-z$ plane of each system. Fig.3a corresponds to an
early snapshot $t=20T_{hmct}$ far from equilibrium, while Fig.3b
corresponds to $t=300T_{hmct}$, at which all systems have approached
an equilibrium state. We see that, for $t=20T_{hmct}$, all the
experiments present a nearly constant ellipticity along the bar,
while, as the systems relax, the ellipticity profile acquires a
declining slope, the isophotes becoming nearly round, i.e. the 3D
matter distribution becoming nearly oblate beyond the end of the
bar. The calculations for $t=20T_{hmct}$ (Fig.3a) are limited within
a radius $R=1r_{hm}$ because all the experiments, except QR1,
present conspicuous spiral structures beyond this radius at
$t=20T_{hmct}$ so that the isophotes in this region cannot
approximated by ellipses.

Boxiness of the isophotes is a well known feature in barred
galaxies. This effect can be measured by the shape parameter $c$,
which can be determined by the equation of the generalized ellipse
(Athanassoula et al. 1990):
\begin {equation}\label{cparam}
\left(\frac{|y|}{b}\right)^c+\left(\frac{|a|}{b}\right)^c=1
\end{equation}
where $a$ and $b$ are the major and minor semi-axes of an isophote
and $c$ is the parameter describing the shape of a generalized
ellipse best-fitting the isophote. For $c=2$ we obtain a standard
ellipse, for $c>2$ the shape approaches a rectangular parallelogram
and for $c<2$ we have a shape similar to lozenge.
\begin{table*}
\caption{Orbital content of the four $N$-body experiments at two
snapshots.} \centering \label{tabt150}
\begin{tabular}{@{}lcccccccc@{}}\hline 
  experiments & QR1 & QR1 & QR2 & QR2 & QR3 & QR3 & QR4 & QR4 \\ \hline
  snapshots & $t=20$ & $t=300$ & $t=20$ & $t=300$ & $t=20$ & $t=300$ & $t=20$ & $t=300$ \\ \hline
  $\Omega_p (rad/T_{hmct})$ & 0.4 & 0.36 & 0.7 & 0.46 & 0.82 & 0.52 & 0.96 & 0.59 \\ \hline
  \% chaotic motion & 59.6 & 57.9 & 61.8 & 57.3 & 63.0 & 59.4 & 56.6 & 56.4\\
  \% chaotic motion inside corotation & 28.5 & 27.8 & 12.6 & 16.0 & 9.8 & 14.9 & 7.4 & 11.1\\
  \% regular motion (2:1) & 12.6 & 16.5 & 18.4 & 21.9 & 14.9 & 17.8 & 21.6 & 21.2\\
  \% regular motion (3:1) & - & - & 18.7 & - & 19.8 & - & 18.5 & -\\
  \% regular motion (A) & 25.5 & 20.4 & -& 15.8 & - & 21.5 & -& 20.8\\
  \% regular motion (B) & -& 4.0 & -& 2.6 & - & 1.1 & - & 1.2\\
  \% chaotic motion (2:1) & 1.8 & 8.7 & 1.7 & 2.7 & 1.3 & 2.0 & 1.1 & 1.7\\
  \% chaotic motion (3:1) & 5.4 & 4.7 & 5.6 & 3.7 & 5.0 & 3.7 & 4.5 & 3.0\\
  \% chaotic motion (4:1) & 4.2 & 6.5 & 4.3 & 4.4 & 4.2 & 1.5 & 3.1 & 2.7\\
  \% chaotic motion (A) & 4.8 & 12.4 & - &9.7 & - & 7.9 & - & 6.0 \\
  \% chaotic motion (cor) & 6.2 & 6.4 & 2.9 & 2.6 & 2.5 & 1.7 & 2.6 & 2.1\\
  \% chaotic motion (-2:1) & 2.4 & 3.2 & 2.4 & 1.7 & 2.5 & 0.7 & 3.1 & 0.6\\
  \% chaotic motion (-1:1) & - & 2.0 & - & 2.3 & 4.4 & 1.9 & 4.4 & 2.1\\
  \% chaotic motion ($q<-1$) & 14.8 & 8.7 & 30.0 & 23.1 & 35.2 & 33.4 & 26.1 &
  35.7\\ \hline
\end{tabular}
\end{table*}

Fig.4 shows the shape parameter $c$ as a function of the pattern
speed of the bar at the time snapshots specified within the figure
for each experiment. Each point's ordinate corresponds to the mean
value of the parameter $c$ of the generalized ellipses having major
semi-axis between $0.4r_c$ and $0.7r_c$, where $r_c$ is the system's
corotation radius at the indicated snapshot. The abscissa of each
point corresponds to the associated pattern speed of the bar. For
values of $\Omega_p$ smaller that $\approx0.7 rad/T_{hmct}$ the
corresponding $c$-parameter seems to have a flat behavior with
$c>2$. However, for values greater than $\approx0.7 rad/T_{hmct}$ a
declining relation is apparent indicating that the boxiness
decreases as the pattern speed $\Omega_p$ increases. Similar study
for elliptical galaxies has been presented in \citet{b53,b52}.

\begin{figure}
\centering
\includegraphics[width=8.1cm]
{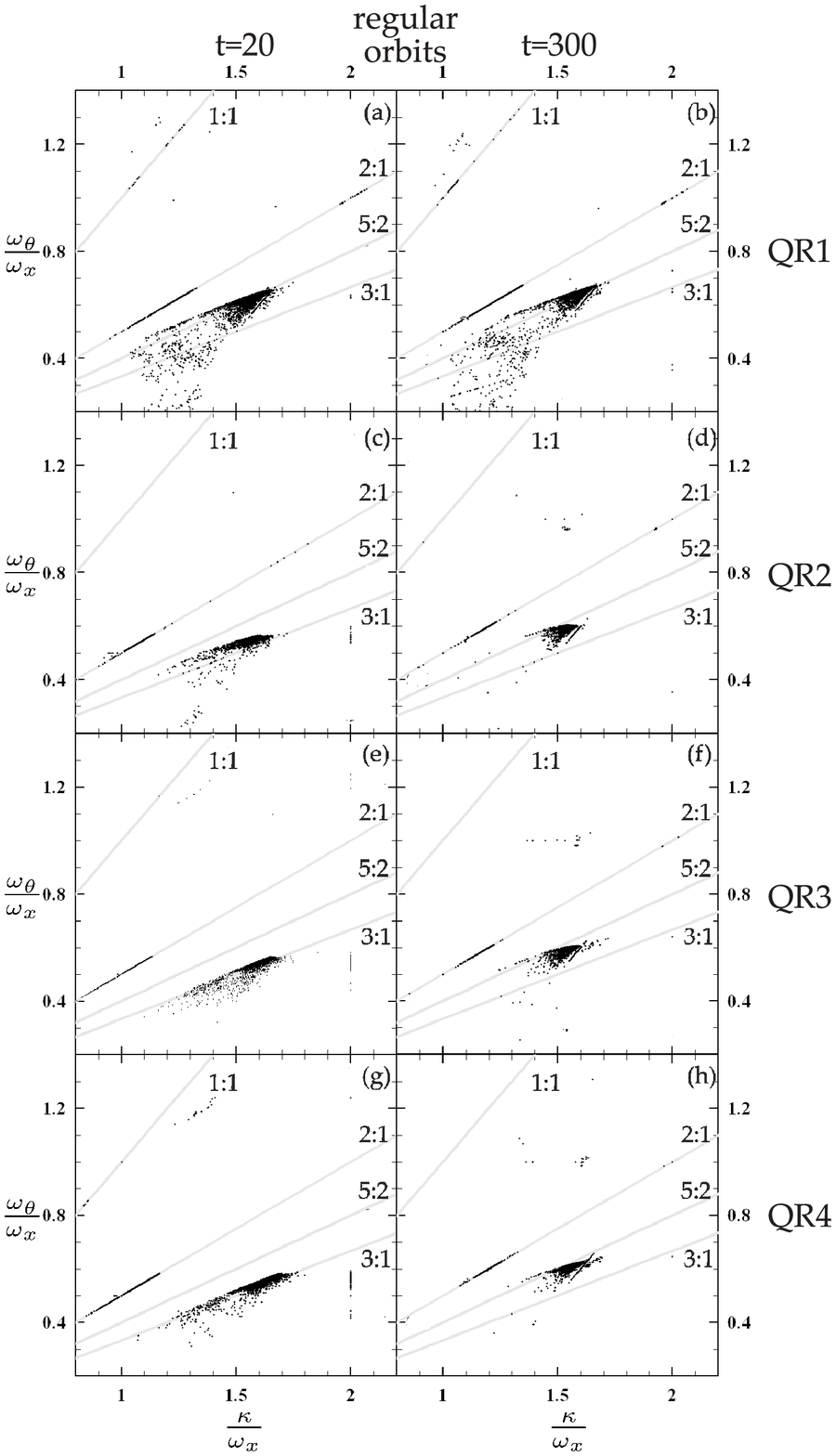} \caption{The frequency map for the regular orbits at
$20T_{hmct}$ (left column) and at $300T_{hmct}$ (right column) for
the four experiments. The `disc' resonances of the form
$m_1\frac{\omega_\theta}{\omega_x}+m_2\frac{\kappa}{\omega_x}=0$ are
shown by gray lines.} \label{fig:5}
\end{figure}

\section{Study of the orbital structure}

\subsection{Frequency analysis of the orbits}

The distinction of the regular from the chaotic orbits is based on a
combination of two methods, namely the {\it Specific Finite Time
Lyapunov Characteristic Number} (SFTLCN) or simply $L_j$, introduced
by \citet{b31}, and the {\it Smaller ALignment Index}, (SALI), or
simply Alignment Index ($AI_j$) \citep{sk2001,b30}. For more details
on the implementation and efficiency of these combined methods see
Paper I.

The percentage of chaotic orbits in all the experiments is close
to $60\%$ (see Table 1), which is almost twice the corresponding
percentage found in the non-rotating progenitor model, i.e. the
Q-model \citep{b30}. The general conclusion in \citet{b31} is that
rotation enhances chaos, not only as regards the fraction of mass
in chaotic motion, but also as regards the magnitude of the
Lyapunov numbers that are seriously shifted towards higher values.
The authors argued that rotation introduces instabilities and a
sequence of important overlapping resonances between the pattern
speed $\Omega_p$ and the frequencies of orbits. This fact is
imprinted in the increase of the chaotic motion. The same
conclusion has been reached in other studies dealing with orbits
in bar like galaxies. For example, \citet{b170}, noted that an
increase of the pattern speed of the bar model simulating the
Milky way increases significantly the chaotic regions on the
Poincar\'{e} map because of resonance overlap phenomena. Also,
\citet{b180}, found that the combination of a triaxial halo with a
fast-rotating bar leads to the development of chaos. Finally,
\citet{b160} found that the fraction of the chaotic orbits is
enhanced in triaxial stellar systems when rotation is inserted.
However, in all the experiments the fraction of mass that can
develop effective chaotic diffusion within a Hubble time is less
than $45\%$ \citep[fig.12b in][]{b31} which means that a
significant percentage of orbits are weakly chaotic.

In the experiment QR1 there appears a transient trailing spiral
structure as a result of the transfer of angular momentum to the
material outwards \citep{b11,b12,b13,b14}. The transient spiral
arms disappear quickly (after $\approx15 T_{hmct}$). As a result,
the system resembles a bar-like galaxy without any spiral
structure. The orbital analysis of this experiment at
$t=300T_{hmct}$ was the subject of Paper I.

Here, we exploit the tool of frequency analysis \citep{b2,b60,b59}
using the improved code of \citet{b48}. We apply this technique
separately for the regular and the chaotic orbits in all four
experiments and for two different snapshots corresponding to
20$T_{hmct}$ and 300$T_{hmct}$, respectively. In order to obtain the
frequency analysis of the orbits at one snapshot, all the orbits are
integrated forward in the `frozen' potential corresponding to that
snapshot, and for a time equal to 150 radial periods. The Jacobi
constant $E_j$ of one orbit is given by the relation
\begin{equation}
E_j=\frac{1}{2}(v_x^2+v_y^2+v_z^2)+V(x,y,z)-\frac{1}{2} \Omega_p^2
R_{yz}^2
\end{equation}
where $V(x,y,z)$ is the full 3D `frozen' potential, given by the
SFC code as an expansion of a bi-orthogonal basis set, $v_x, v_y,
v_z$ are the velocities in the rotating frame of reference and
$\Omega_p$ is the angular velocity of the bar at the studied
snapshot (see Table 1 for the values of $\Omega_p$ for all the
models in the two snapshots). The value $R_{yz}$ is the distance
from the rotation axis.

The integrated orbits belong in a wide range of Jacobi integral
values which implies a wide range of orbital periods (e.g. radial
periods) of each orbit. This means that within a certain time (e.g
a Hubble time) each orbit has a quite different dynamical
evolution. In order to treat this integration process with the
minimum computational cost, we have used a Runge-Kutta 7(8) order
integrator with variable time step. This technique guaranties a
relative error of the Jacobi constant less than $10^{-8}$ for all
the integrations we provide below. Moreover, by requiring a
relative error less than $10^{-9}$ we obtained very similar
results, which implies that the presented results are robust.

By running the orbits in the `frozen' potential, we then calculate
the following frequencies for each orbit:

a) the radial frequency $\kappa$ (frequency of epicyclic
oscillations) which is derived from the time series of the radial
velocity $\dot{r}(t)$

b) the angular frequency in the rotating frame of reference,
$\omega_{\theta}$, or the inertial frame, $\Omega$, which are
derived from the time series of the corresponding polar angle

c) the vertical frequency, $\omega_x$ which is derived from the time
series of the coordinate $x(t)$.

\begin{figure}
\centering
\includegraphics[width=8.1cm]
{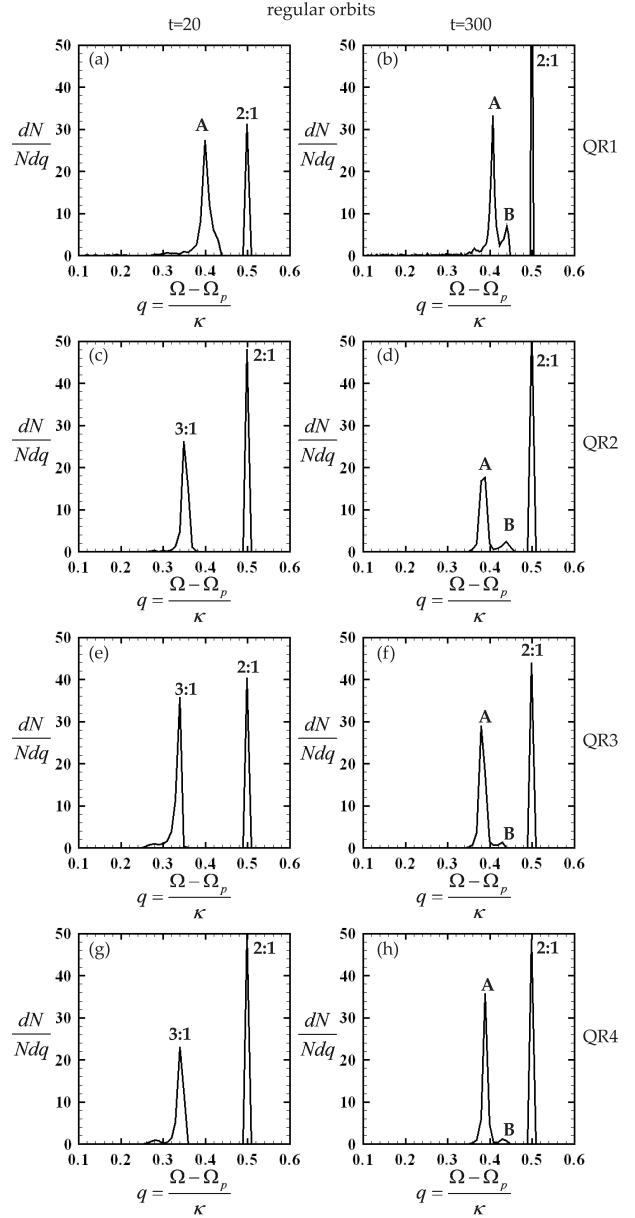} \caption{The distribution of the frequency ratio $q$
for the regular orbits at 20$T_{hmct}$ (left column) and at
300$T_{hmct}$ (right column). The main type of orbits supporting
the bar is the 2:1 resonant orbit (or `x1' type of orbits, using
the nomenclature of Contopoulos \& Papayannopoulos, 1980) for both
snapshots. The second important type of orbits is either the 3:1
resonant orbit or the group `A' depending on the system and on the
time.} \label{fig:6}
\end{figure}

The main resonances can now be detected using frequency maps.
Figure 5 shows the rotation numbers
$(\frac{\kappa}{\omega_x},\frac{\omega_{\theta}}{\omega_x})$ of
the ensemble of {\it regular} orbits for all the experiments at
the two studied snapshots. Each row (column) of panels corresponds
to a different system (snapshot) as indicated in the figure. In
each panel, one point corresponds to one orbit of a real $N$-body
particle that turns to be regular. In the `frozen potential'
approximation the position of this point on the
$(\frac{\kappa}{\omega_x},\frac{\omega_{\theta}}{\omega_x})$ plane
remains invariant in time, since the particle's orbit stays
confined on an invariant torus with the given values of the
rotation numbers.

The most important resonances are the `disc' resonances which are of
the form
$m_1\frac{\omega_\theta}{\omega_x}+m_2\frac{\kappa}{\omega_x}=0$,
with $m_1,m_2$ integers.  Such resonances are shown by gray straight
lines in Fig.5. Orbits lying exactly on them belong to two groups:

\begin{figure}
\centering
\includegraphics[width=7cm]
{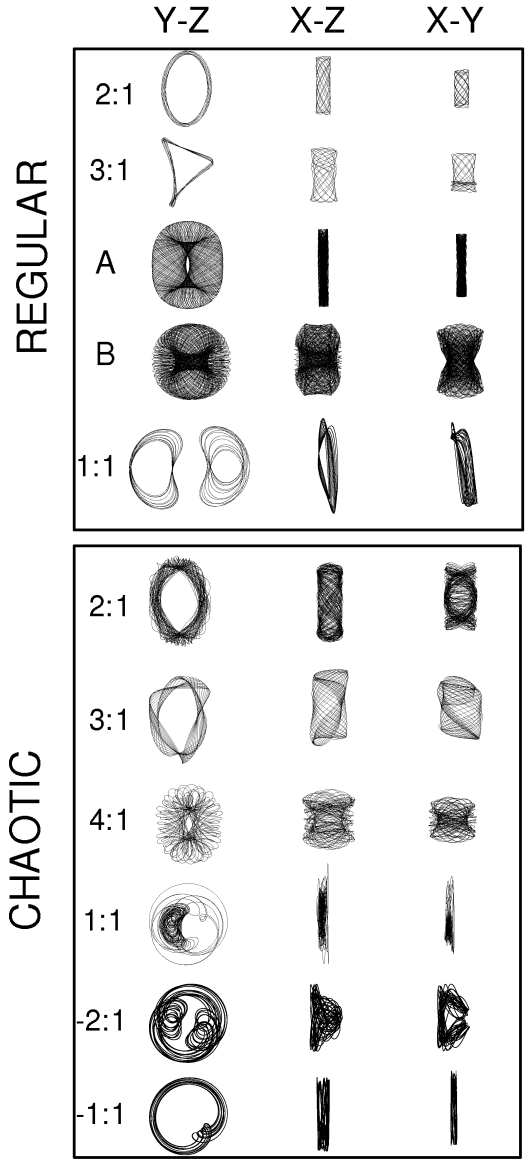} \caption{The three projections of real $N$-body orbits
corresponding to the most important resonances. Regular orbits are
plotted in the upper panel and chaotic ones in the lower panel.
Note that the bar's major axis is along the $z$-axis.}
\label{fig:7}
\end{figure}

a1) if no second resonance relation of the form
$m_1'\frac{\omega_\theta}{\omega_x}+
m_2'\frac{\kappa}{\omega_x}+m_3'=0$, with all three integers
$m_i'\neq 0$, is satisfied, the orbit lies on a torus of dimension
smaller than or equal to three. The orbit is quasi-periodic, and it
is the product of a vertical oscillation and of a motion on the
$y-z$ plane which is either exactly periodic or a `tube' around a
planar periodic orbit. The motion in the third dimension is
essentially decoupled from the motion on the 2D plane. In practice,
we find that the great majority of $N$-body particles in exactly
resonant orbits populate precisely the above group. Furthermore,
most of them exhibit vertical oscillations of a rather small
amplitude, compared to the motion on the $y-z$ plane (this is
expected since the system as a whole is substantially flattened in
the $x$-axis). Thus a study of dynamics via a 2D approximation is
sufficient to unravel all interesting properties of such orbits.

a2) if a second resonance relation of the form
$m_1'\frac{\omega_\theta}{\omega_x}+
m_2'\frac{\kappa}{\omega_x}+m_3'=0$ is satisfied, the orbits are
called `doubly resonant'. The orbits can be exactly 3D-periodic,
corresponding e.g. to a vertical bifurcation from the x1 family
(see Skokos et. al 2002 for a classification and nomenclature of
such periodic orbits), or thin tubes around these periodic orbits.
Particles on such orbits influence mainly the `edge-on' profiles
of our systems. The study of these orbits is outside the scope of
the present paper and will be undertaken in a separate study. The
connection between the `disc' resonances and the orbital
frequencies is explicitly defined, in Athanassoula (2003).

On the other hand, particles satisfying a `disc' resonance relation
only approximately, i.e.
\begin{equation}\label{apres}
m_1\frac{\omega_\theta}{\omega_x}+m_2\frac{\kappa}{\omega_x}\approx 0
\end{equation}
can  also be distinguished in two cases:

b1) if there is one exact resonance relation of the form
$m_1'\frac{\omega_\theta}{\omega_x}+m_2'\frac{\kappa}{\omega_x}+m_3'=0$,
$m_i'\neq 0$, the orbit lies on an invariant 2D torus. When
projected on the $y-z$ plane, however, the orbit is {\it not}
captured in the associated $m_1:-m_2$ disc resonance, i.e., it lies
entirely outside the separatrix domain corresponding to that
particular resonance under a 2D approximation of the dynamics. Such
orbits exhibit a slow precession around the associated exact disc
resonances and they are responsible for a number of interesting
features of the `face on' appearance of the galaxies, examined in
detail below.

b2) if there is no exact resonance relation of the form
$m_1'\frac{\omega_\theta}
{\omega_x}+m_2'\frac{\kappa}{\omega_x}+m_3'=0$ the orbit is
quasi-periodic, lying on a 3D torus. When projected on the $y-z$
plane, these orbits share most features of the orbits of group (b1).

In Fig.5, we observe that the majority of particles in regular
orbits are concentrated along and around specific resonance lines. A
very clear concentration is seen along the resonance 2:1 (that
corresponds to $m_1=2$ and $m_2=-1$ in Eq.4). These are
quasi-periodic orbits forming thin tubes around the `x1' type of
periodic orbits \citep{b47} at values of the Jacobi constant
corresponding to distances near the inner Lindblad resonance
(hereafter ILR). Both the periodic and quasi-periodic orbits of this
type have ellipsoidal forms in the 3D space, which are projected to
elliptical figures in the $y-z$ plane with a long axis aligned to
the long axis of the bar (Fig.7).

The 2:1 is the main resonance supporting the bar and therefore it is
well populated in all four experiments both at 20$T_{hmct}$ and at
300$T_{hmct}$. On the other hand, depending on the system and
snapshot considered, other significant groups of regular orbits
supporting the bar are distinguishable. We note that the simple
resonances of the form $n:1$, $n>0$, implying $n$ epicyclic per one
azimuthal oscillation, are concentrated, as $n$ increases, to
domains of the phase space (or values of the Jacobi constant) closer
and closer to corotation. Since the accumulation of these resonances
causes chaos by the mechanism of resonance overlap (see Contopoulos
2002 p.185), the regular orbits are expected to lie at resonances
$n:1$ with a low value of $n$, or some nearby non-simple resonances
$n':m' \simeq n:1$, while the chaotic orbits fill the remaining
parts of the resonance web, i.e. the simple resonances $n:1$ with
high values of $n$ or their nearby non-simple resonances
$n':m'\simeq n:1$.

\begin{figure*}
\centering
\includegraphics[width=16cm]
{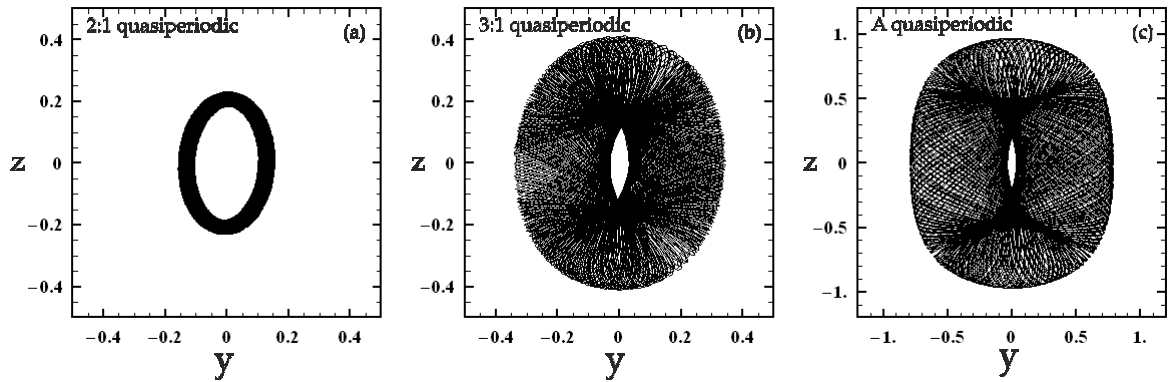} \caption{a) a 2:1 resonant orbit b) a 3:1 resonant
orbit and c) an orbit of group `A'. Note that the bar's major axis
is along the $z$-axis. All orbits are integrated for a time
comparable to the Hubble time.} \label{fig:8}
\end{figure*}

Figure 6 gives the distribution of regular orbits along the quantity
$q=\frac{\Omega-\Omega_p}{\kappa}$ calculated for all the systems at
20$T_{hmct}$ (left hand column) and at 300$T_{hmct}$ (right hand
column). Note that the pattern speed $\Omega_p$ of the bar at any
snapshot can be derived from fig.4a of \citet{b31}. It is well known
that the quantity $q$ distinguishes very efficiently the various
resonances and it has been used for the orbital classification in
previous studies \citep[e.g.][]{b14,b16}.

Each peak in Fig.6 corresponds to the indicated population.
Besides the ILR (at $q=\frac{1}{2}$) we identify other important
peaks in the $q$-value distribution of the $N$-Body particles. One
prominent peak which appears at early snapshots (except in the
experiment QR1) is at the 3:1 resonance (that corresponds to
$m_1=3$ and $m_2=-1$ in Eq.(4) and to $q=1/3$ in Fig.6). This peak
nearly disappears at late snapshots, being replaced, instead, by a
different peak, called hereafter group `A'. This corresponds to
particles with a $q$-value $q\simeq\frac{2}{5}$. This population
has been found in other $N$-body simulations of bar-like galaxies
\citep[see for example][]{b17,b16}. The transition from a
distribution peaked at the 3:1 resonance to a distribution peaked
at the group `A' is related to a morphological transition in the
bar, discussed in detail below.

A less populated group called group `B', is also observed at
snapshots near equilibrium (e.g. $t=300T_{hmct}$). This consists
mainly of quasi-periodic retrograde orbits with a $y-z$ projection
of the `x4' type (Contopoulos \& Papayannopoulos 1980), i.e.
elongated perpendicularly to the bar and correspond to $m_1=2$ and
$m_2=-1$ in Eq.(4). Remarkably, many orbits of this group exhibit
also vertical oscillations (see Fig.7). The group `B' orbits
cannot exist at snapshots close to the beginning of the
simulation, as for example $t=20T_{hmct}$, since the `velocity
re-orientation' process, by which the initial conditions are
produced, implies that there be no retrograde orbits initially.
Thus the group `B' population develops in the course of the
simulation, as the systems evolve towards the equilibrium. From
the right panels of Fig.6 we can conclude that the number of
particles in the group `B' decreases with increasing pattern speed
(e.g. from QR1 to QR4). This is further substantiated in section
3.2 in which we show that in all the self-consistent systems the
phase space corresponding to the `x4' family is almost empty.

Finally, in the experiment QR1, which does not have any spiral
structure, there is a small concentration of particles in regular
orbits along the resonance line 1:1 (see Fig.5). This resonance
corresponds to the short period orbits PL4 and PL5 (using the
nomenclature of Voglis et al. 2006b for the short period periodic
orbits, bifurcating from the corresponding Lagrangian points),
around the Lagrangian points $L_4$ and $L_5$ (see Fig.7) and
exists for values of the Jacobi constant close to the one at
corotation. In all the other experiments there are no regular
orbits of this type.

The implications of the transition from a 3:1 peak at early
snapshots to a group `A' peak at late snapshots can be discussed
with the help of Figs.7 and 8. Figure 7 shows the projections on
various planes of some regular orbits trapped in major disc
resonances. It is obvious that the group `A' orbits yield a `boxy'
form, while the resonant 3:1 orbits are asymmetric with respect to
the bar's major axis (we can find a different orbit, symmetric to
the 3:1 orbit of Fig.7, with respect to the axis of symmetry $y=0$).
However, other orbits of the $q=\frac{1}{3}$ group of Fig.6, which
are selected so that they are not exactly resonant (i.e. $q\simeq
\frac{1}{3}$), exhibit a precession of the apocentres which
corresponds to a motion in the phase space close to, but outside the
separatrices marking the 3:1 resonant domain. Such an orbit is shown
in Fig.8b. Orbits such as in Fig.8b are considerably less boxy than
the orbits of group `A' (Fig.8c), although they are more boxy than
the 2:1 resonant orbits (Fig.8a). We conclude that if the
$q=\frac{1}{3}$ peak of the distribution is pronounced compared to
other peaks, the bar isophotes appear more `disky', i.e lower $c$
values (see Fig.4 for the early snapshots of the experiments QR2 to
QR4). On the contrary, the enhancement of group `A' orbits is
responsible for the `boxy' isophotes (high $c$ values in Fig.4) at
the late snapshots of the same experiments. This implies also a
correlation between the position of the peaks of the
$q$-distribution and the value of the pattern speed of a particular
galaxy. The 3:1 type of orbits appears mainly in systems with a high
value of $\Omega_p$ (e.g. the systems QR2, QR3, QR4 at
$t=20_T{hmct}$), while, as the systems slow down, the $q=1/3$ peak
becomes gradually depleted and, instead, the $N$-body particles
gradually populate the group `A' type of orbits, which, hence,
becomes a dominant type in systems with a low value of $\Omega_p$
(QR2, QR3, QR4 at $t=300T_{hmct}$). Note that the pattern speed of
the QR1 experiment is quite small already at $t=20T_{hmct}$, and, in
this particular experiment, the group `A' peak is prominent also at
this snapshot. This discussion suggests that $\Omega_p$ is indeed,
the relevant parameter to which the position of the main peaks of
the $q$-distribution should be correlated.

\begin{figure}
\centering
\includegraphics[width=8.1cm]
{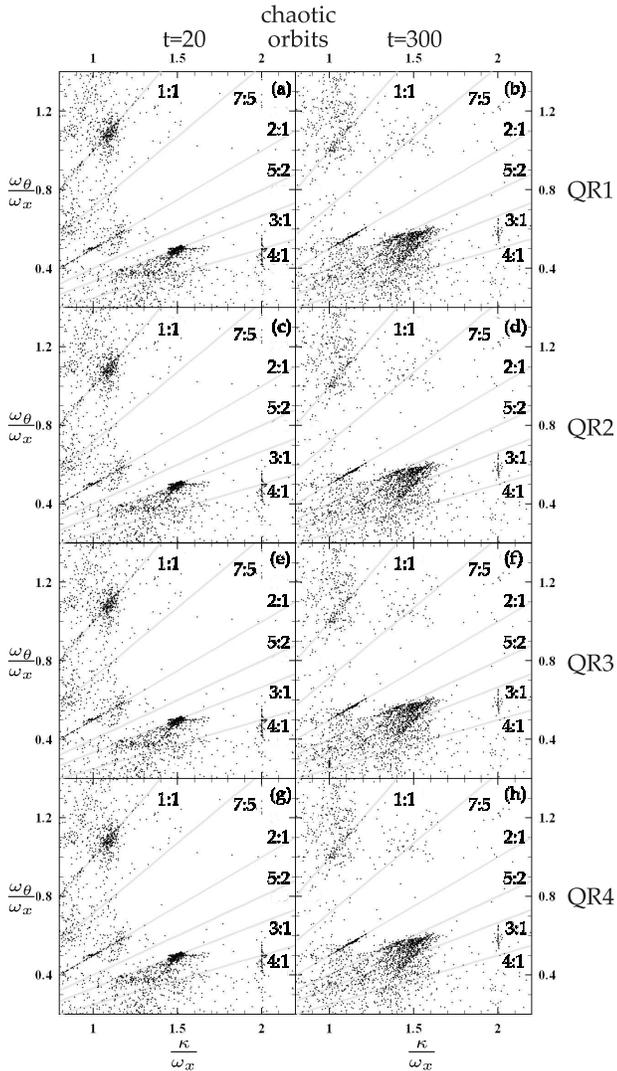} \caption{Same as in Fig.5 but for the chaotic orbits. We
see that chaotic orbits appear quite scattered. Nevertheless, many
chaotic orbits are trapped near various resonance lines in some of
which no regular orbits are observed.} \label{fig:9}
\end{figure}

Figure 9 shows the frequency maps of the {\it chaotic} orbits, for
all the systems and for two different snapshots as indicated in the
figure. It is obvious that the chaotic orbits appear quite scattered
in these diagrams. Nevertheless, many chaotic orbits are trapped
near various resonance lines. Such orbits are called `sticky' and
they behave similarly to regular orbits for long times. An
interesting remark is that there are chaotic orbits near resonance
lines where no regular orbits are observed. For example, in Fig.9 we
can see points around the 3:1 and the 4:1 resonance as well as
around the 1:1 resonance (see Fig.7 for their shapes), at snapshots
when there are only a few regular orbits in these resonances
(compare Figs. 5 and 9).

\begin{figure}
\centering
\includegraphics[width=8.1cm]
{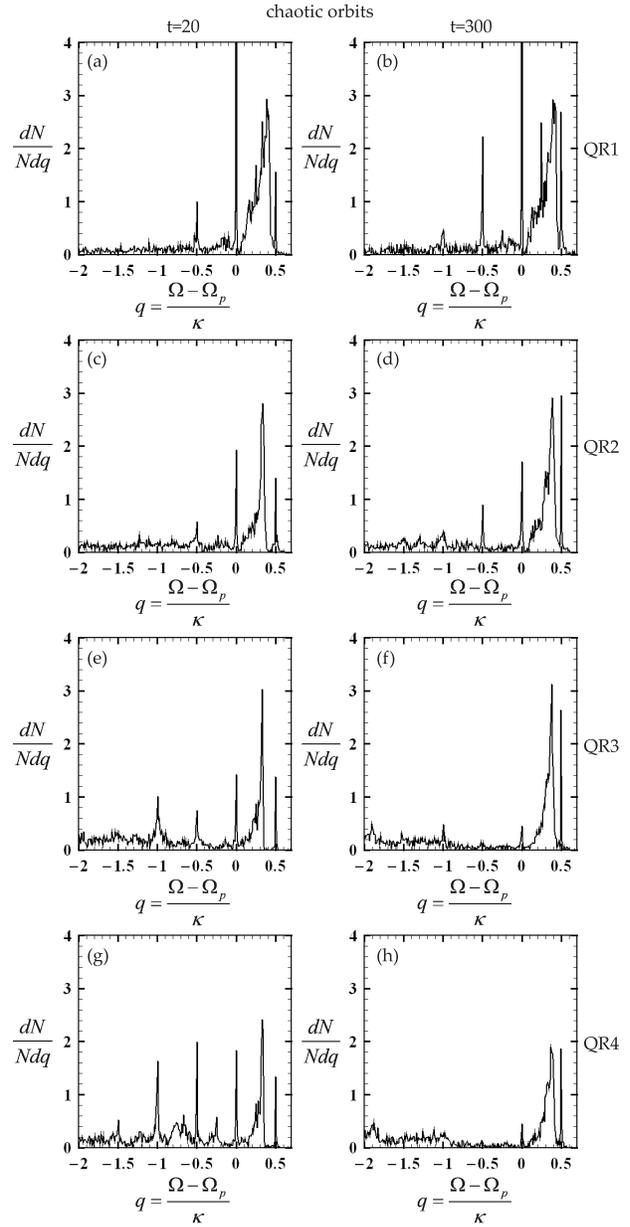} \caption{The distribution of the frequency ratio $q$ for
the chaotic orbits at $20T_{hmct}$ (left column) and at
$300T_{hmct}$ (right column). Orbits inside corotation have positive
$q$ values, while orbits extending outside corotation have negative
$q$ values. The chaotic orbits are distributed in a greater variety
of $q$ than the regular ones.} \label{fig:10}
\end{figure}

Figure 10 shows the distribution of the $q$ values for the chaotic
orbits of all the systems. The first remark is that the chaotic
orbits are distributed in a greater variety of $q$-values than the
regular ones. The $q$-distributions present peaks at the same
values of $q$ as for the regular orbits. Nevertheless, important
peaks appear also at other resonances, as seen in the figure. Note
that corotation, ILR and outer Lindblad resonance (hereafter OLR)
correspond to $0,\frac{1}{2}$ and $-\frac{1}{2}$ values of $q$,
respectively. In general orbits located inside (outside)
corotation have positive (negative) $q$ values. The orbits
corresponding to smaller $q$ values reach larger distances on the
$y-z$ plane. Since at $t=300T_{hmct}$ all the systems have lower
$\Omega_p$ values than at $t=20T_{hmct}$, the corotation radius is
at larger distances. This explains the smaller percentage of
orbits found outside corotation ($q \leq 0$) at $t=300T_{hmct}$.

The percentages of the most important populations of regular and
chaotic orbits, derived from Figs.6 and 10, are given in Table 1.

\begin{figure*}
\centering
\includegraphics[width=16cm]
{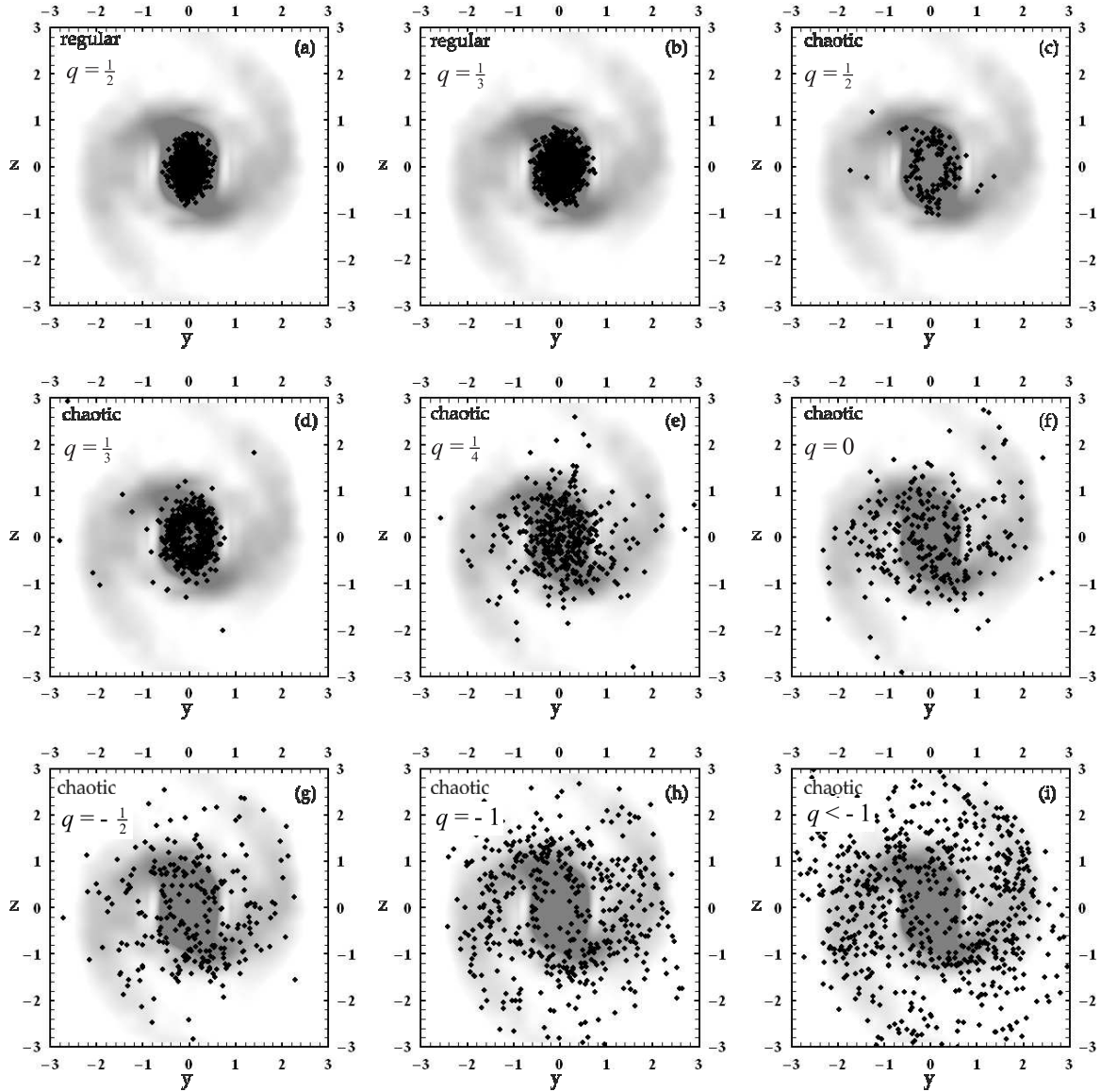} \caption{The instantaneous positions of the particles
belonging to the different populations of the regular and chaotic
orbits (black dots) superimposed on the backbone of the galaxy on
the plane of rotation (gray background) for the experiment QR3 at
$t=20T_{hmct}$.} \label{fig:11}
\end{figure*}

We now come to see the role of the various populations in supporting
particular morphological features of the studied systems. In
subsequent plots we focus on one example, the QR3 experiment, which
exhibits all interesting phenomena. The `backbone' of the system,
i.e. its appearance on the $y-z$ plane, is plotted as a gray
background in a number of subsequent plots.

Figure 11 shows the `backbone' of the system QR3 at $t=20T_{hmct}$
together with the instantaneous positions of the particles belonging
to the various identified populations of regular and chaotic orbits.
The backbones of QR2, QR3 and QR4 are similar at $t=20T_{hmct}$. All
four experiments present a bar, two spiral arms and a faint ring. We
observe that the regular orbits contribute to the form of the bar.
On the contrary, the chaotic orbits support the structures both
inside and outside corotation. The chaotic orbits inside corotation
($q \ge 0$) create an envelope of the bar, the particles being
located at the outer bar layers. Similar orbits were found in
\citet{b301}. The particles in chaotic orbits near the 4:1 resonance
($q=\frac{1}{4}$) support segments of the spiral arms connected to
the bar. The population corresponding to corotation $(q=0)$
contributes to the outer bar as well as to the ring. The chaotic
orbits located near the OLR $\left(q=-\frac{1}{2}\right)$, as well
as near the $-1:1$ resonance $\left(q=-1\right)$, contribute to the
ring and to the spiral arms. The chaotic orbits below the $-1:1$
resonance $\left(q \le -1\right)$ contribute mainly to the spiral
arms.

\begin{figure*}
\centering
\includegraphics[width=16cm]
{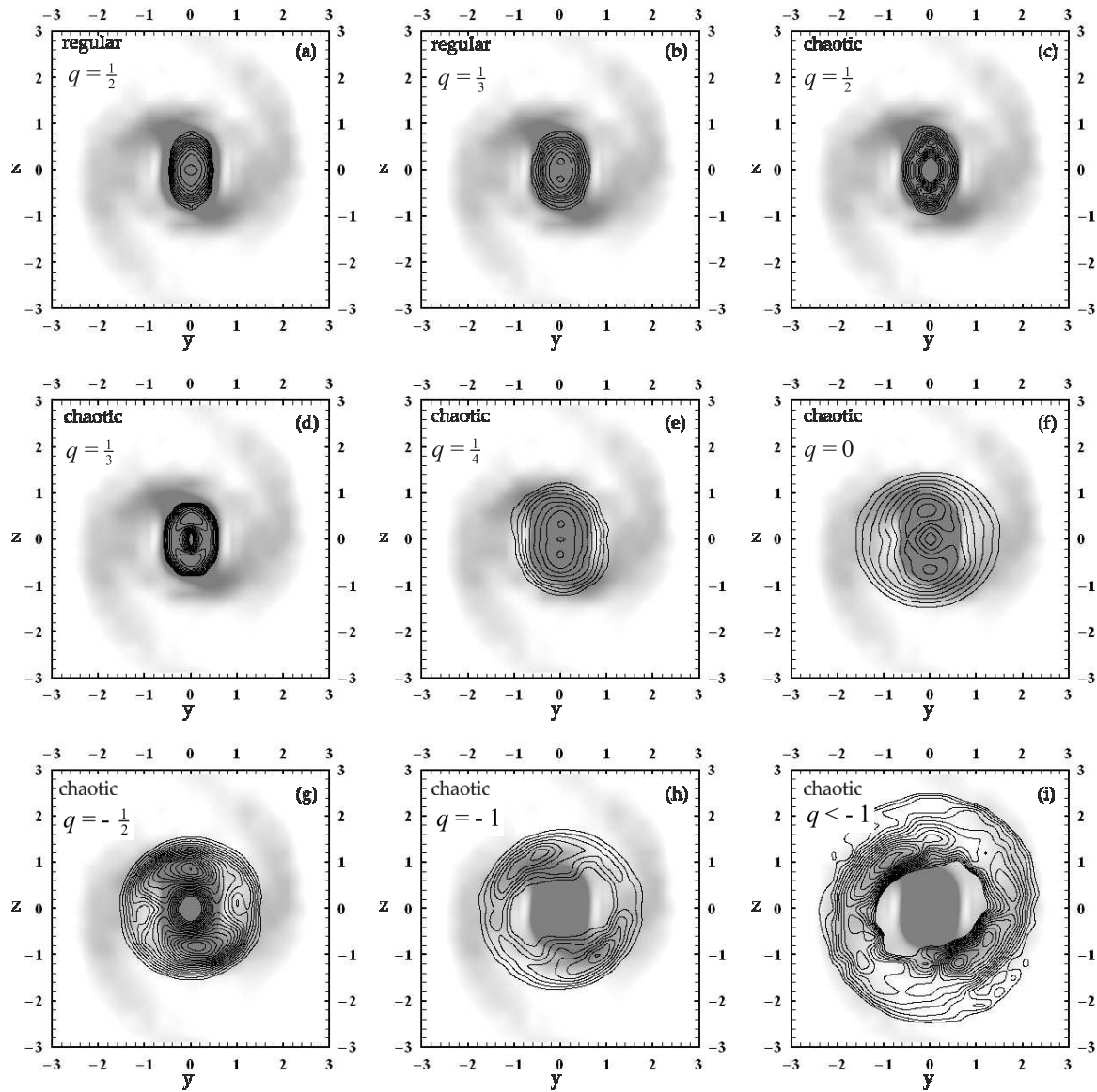} \caption{The isodensities of the different populations
of regular and chaotic orbits derived from their integration for
$400T_{hmct}$ (black lines), superimposed on the backbone of the
galaxy on the plane of rotation (gray) for the experiment QR3 at
$t=20T_{hmct}$.} \label{fig:12}
\end{figure*}

In Fig.11  we only see which features of the system are supported by
the instantaneous positions of the particles. Thus we have no
information about the trend for morphological changes induced by the
chaotic orbits undergoing slow chaotic diffusion. In order to see
such effects, Fig.12 shows again the backbone of the system QR3 at
$t=20T_{hmct}$ together with the isodensities of the high density
areas (black contours) that we get from the superposition of the
orbits of the same particles as in Fig.11, integrated for a time
interval $400T_{hmct}$ (a little longer than a Hubble time). The
isodensities are plotted separately for each population, while the
numerical integration is carried under the fixed gravitational
potential of the system at $t=20T_{hmct}$. We now see very clearly
which features of the system are supported by each type of orbits in
the long run. An important remark is that the chaotic orbits outside
corotation $\left(q \le 0\right)$, which can in principle escape,
continue in practice to support the basic features of the system
after an integration time comparable to the Hubble time. In
particular, the populations with $q=0$ and $q=-\frac{1}{2}$ support
the ring around the bar, the population with $q=-1$ supports the
spiral arm segments connected to the end of the bar and the
population with $q \leq -1$, supports the outer parts of the spiral
arms, as well. As shown in Section 3.2, these orbits have higher
values of the Lyapunov exponents, implying a faster chaotic
diffusion. The chaotic diffusion is a major factor driving the
secular evolution of the systems within times comparable to the
Hubble time.

\begin{figure}
\centering
\includegraphics[width=8.5cm,angle=0]
{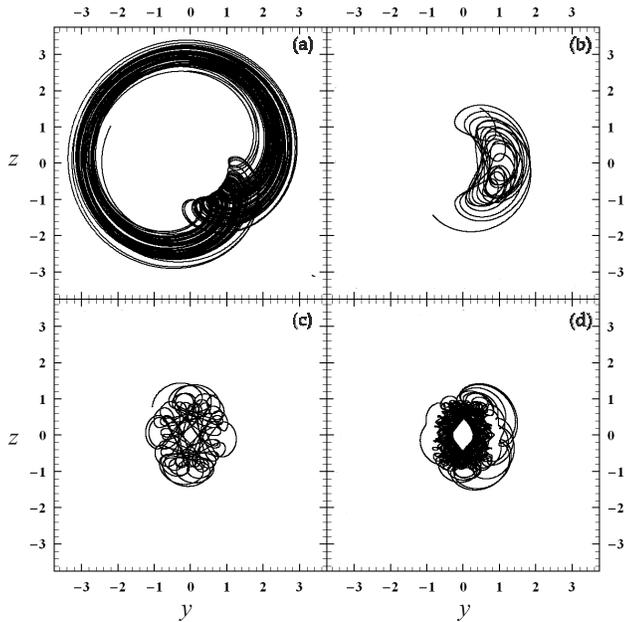} \caption{The evolution of a real chaotic orbit (of the
experiment QR4 at $t=20T_{hmct}$), that presents `stickiness' to
several resonances before escaping from the system after more than a
Hubble time. Different panels correspond to different intervals of
time integration, of the same orbit. The orbit is located
successively around frequency ratios: (a) $q=-1$ (b) $q=0$ (c)
$q=\frac{1}{4}$ and (d) $q=\frac{1}{2}$.} \label{fig:13}
\end{figure}

It should be stressed here that the percentages of the various
types of chaotic orbits change in time as a result of i) the
chaotic diffusion, and ii) the gradual change of the pattern
speed. Thus, orbits of a specific type can be converted to orbits
of another type, or they can escape. At the equilibrium state this
process must be in dynamical equilibrium (the number of orbits
changing type should be equal to the number of orbits joining
every same type). Thus, in equilibrium, such an exchange can only
be due to the chaotic diffusion. The studied systems at
$t=20T_{hmct}$ are far from equilibrium. Therefore, the diffusion
of the orbits, especially those not trapped by some resonance,
produces a strong secular evolution. An example of such a process
is given in Fig.13 where the evolution of the chaotic orbit of a
real particle, of the experiment QR4, is plotted before its final
escape from the system. It is obvious that this chaotic orbit
stays consecutively localized in a sequence of resonances,
spending a considerable time near each resonance, before escaping
finally from the system, after approximately three Hubble times.
In particular, the orbit has a frequency ratio close to -1:1
($q=-1$) up to about one Hubble time (Fig.13a). Then it stays
located around the Lagrangian point `$L_4$' corresponding to a
frequency ratio $q=0$ (Fig.13b). Figure 13c presents the part of
the orbit that gives a frequency ratio close to $q=\frac{1}{4}$
and finally Fig.13d corresponds the part of the orbit having
frequency ratio close to $q=\frac{1}{2}$. The explanation of such
behaviour of chaotic orbits is given using 2D surfaces of sections
(see section 3.2). For certain Jacobi constants the area outside
corotation can communicate with the area inside corotation and
chaotic orbits present long time `stickiness' in unstable
asymptotic curves of several resonances, until they finally escape
from the system following the path carved by these curves.

\begin{figure*}
\centering
\includegraphics*[width=13cm]
{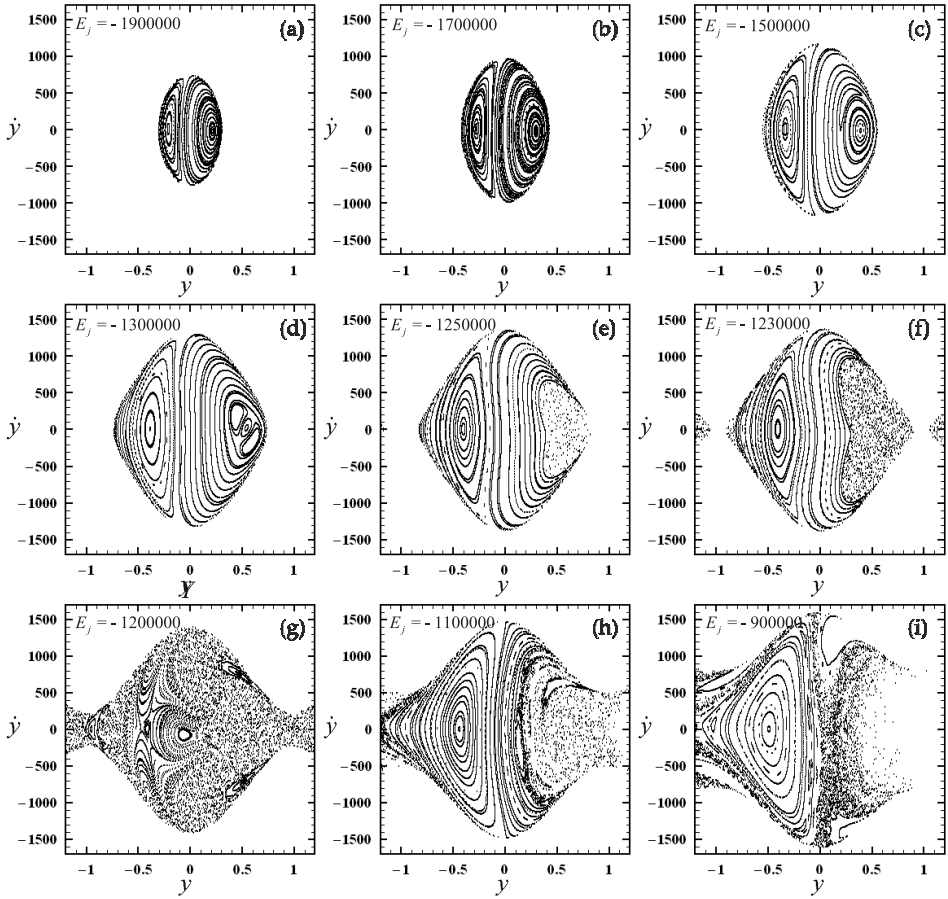} \caption{Surfaces of sections in the 2D approximation
for the experiment QR3 at $t=20T_{hmct}$ and for various Jacobi
constant values.} \label{fig:14}
\end{figure*}

\subsection{Phase space structure}

Interesting remarks about the phase space structures can be revealed
by plotting 2D projections of the 4D surfaces of sections (hereafter SOS)
of all the self-consistent $N$-body models.

Using the `frozen' potential and the instantaneous value of the
pattern speed $\Omega_p$ at each studied snapshot (Table 1) we
construct SOS of test particles for all the systems and thereby
identify the main features of the phase space structure. On the
other hand, plotting the real $N$-Body particles on the same SOS
reveals which domains of the phase space are preferred and which
are avoided by the real particles. A similar technique has been
used in Paper I. A 4D SOS can be constructed by the intersection
of the orbits with the plane $z=0$, $\dot{z}<0$. We then plot the
projection of this particular SOS on the ($y,\dot{y}$) plane.
Alternatively, we can use the 2D approximation of the
gravitational potential on the disc plane and integrate the orbits
of test particles on the same plane. This is hereafter called a
`2D approximation'.

Figure 14 gives the phase space structure at various values of the
Jacobi constant for the system QR3 at t=20$T_{hmct}$, via the 2D
approximation. The central value of the potential (potential well)
and the value of the Jacobi constant at the Lagrangian points
$L_1$, $L_2$ are $E_0=-2.15\times10^6$ and
$E_{L_1}=-1.265\times10^6$ respectively. The panels of the two
upper rows of Fig.14 show the phase space structure for Jacobi
constants below the value corresponding to $L_4, L_5$
$E_{L_4}=-1.225\times10^6$ (Note that the phase space areas
corresponding to inside and outside corotation are divided). It is
obvious that chaos becomes dominant for $E_j>-1.3\times10^6$. The
areas inside corotation with $y<0$ correspond to retrograde orbits
of the `x4' type (and their quasi-periodic orbits), while the
areas with $y>0$ are filled with `x1' orbits and their
bifurcations.

Figure 15 shows the projections of the real $N$-body particles with
energies $E_j=-1.5\times10^6\pm 10^4$ on the SOS. The orbit of each
real particle is integrated for 200 iterations. A distribution of
the particles on layers corresponding to a foliation of invariant
tori is discerned in this figure, despite the projection effects
caused by the fact that the SOS is actually 4D. The central energy
value is the same as in Fig.14c. At this value of the Jacobi
constant there are almost no chaotic orbits inside corotation.
Therefore, the orbits of Fig.15 are regular orbits close to the 2:1
and 3:1 resonances. Note that the area with $y>0$, corresponding to
orbits around the `x1' periodic orbit, is well populated, while the
area with $y<0$ corresponding to the `x4' retrograde type of orbits
is nearly devoid of real particles.

Figure 16 shows the energy (Jacobi constant) distribution of the
real $N$-body particles for the regular orbits (solid lines), and
for the chaotic orbits (dashed lines) for the experiment QR3 at
$t=20T_{hmct}$ (gray) and $t=300T_{hmct}$ (black). From Figs.14 and
16 we conclude that most chaotic orbits, with $E_j\lesssim
-1.3\times10^6$, are located outside corotation (the chaos is
negligible inside corotation). On the other hand, for values greater
than the above threshold there are chaotic orbits found both outside
and inside corotation (mostly at the outer regions of the bar). In
the 2D approximation, the chaotic orbits with energies $E_j\lesssim
E_{L_1}$ which are located inside corotation, (Fig.14) are
surrounded by invariant curves and, consequently, they cannot
communicate with the regions outside corotation. The important
remark is that the above result remains essentially valid for the
orbits in the full 3D potential. Namely, the majority of the real
chaotic orbits of the same energy levels which are located inside
corotation remain there at least up to the end of our numerical
integration, i.e. for times much longer than a Hubble time, although
in the full 3D potential these orbits are in principle able to
escape via the phenomenon of `Arnold diffusion' (Arnold 1964).

\begin{figure}
\centering
\includegraphics[width=6.3cm,angle=0]
{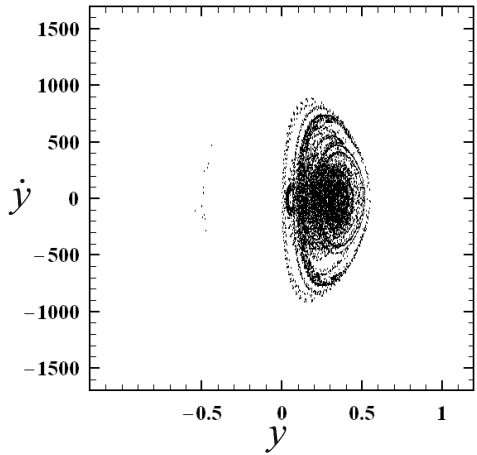} \caption{ (a) The projection on the ($y, \dot{y}$) SOS
of real particles for the region inside corotation for the QR3
experiment, for $Ej=-1.5\times10^6\pm10^4$ at $t=20T_{hmct}$. The
area around `x1' is well populated while the area around `x4'
corresponding to retrograde orbits is almost empty.} \label{fig:15}
\end{figure}

We have found that $\approx 26\%$ of the particles in chaotic orbits
of this experiment are located inside corotation and close to the
end of the bar, therefore contributing to the backbone of the bar.
The percentages of the particles in chaotic orbits which remain
inside corotation, are shown in Table 1, for all the experiments.
From this we conclude that this `chaotic component' represents an
important percentage of particles in all the experiments. The value
of the Specific Finite Time Lyapunov Characteristic Number $L_j$ of
these orbits is in general smaller than the value of $L_j$ of the
chaotic orbits outside corotation. This can be inferred from Fig.17,
in which the $\log(L_j)$ values of the orbits are plotted as
functions of their $q$ values. The chaotic orbits have $L_j\geq
10^{-2.8}$ (see Voglis et al., 2006a). The regular orbits are almost
completely located inside corotation (around the resonances 2:1 and
3:1) and have the smallest values of $L_j$. The chaotic orbits, on
the other hand, are spread in the whole range of resonances, but it
is obvious that the most weakly chaotic orbits (having smaller
$L_j$) are located inside corotation and therefore support the bar.

\begin{figure}
\centering
\includegraphics*[width=7.3cm]
{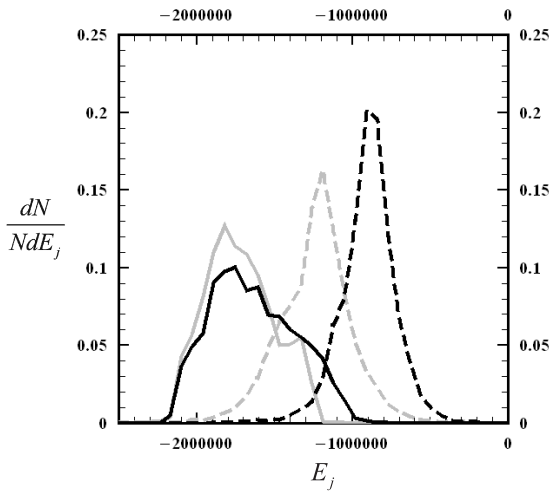} \caption{The energy distribution for regular orbits
(solid lines) and chaotic orbits (dashed lines) for the experiment
QR3 at $t=20T_{hmct}$ (gray) and at $t=300T_{hmct}$ (black).}
\label{fig:16}
\end{figure}

Figure 18a shows another SOS, obtained from the 2D approximation,
for the experiment QR3 at $t=20T_{hmct}$ and for the Jacobi constant
$E_j=-1.1\times10^6$. In this figure we see also the area outside
corotation (which is located at $y\simeq 1.2$). A sticky region is
apparent here between corotation and $y\simeq 2$. This stickiness
phenomenon is related to the unstable manifolds emanating from
unstable periodic points with fixed points in the same area. This
phenomenon, called `stickiness in chaos' has been studied thoroughly
in simple dynamical systems \citep[see][and references
therein]{b15}. The corresponding Jacobi constant is the maximum of
the energy distribution of the chaotic particles having the same
angular velocity as the pattern speed of the bar ($q=0$). The two
stable periodic orbits located inside the sticky zone correspond to
bifurcations of the main orbit around the point $L_5$ (named `PL5')
which has become unstable at this value of Jacobi constant. All the
unstable asymptotic curves are necessarily parallel to the
asymptotic curve of the simplest periodic orbit i.e. the short
period orbit around $L_5$. Thus this orbit is called `PL5'. In
Fig.18b we plot the projection on the same SOS of the real
particles. We observe again the depopulation of the `x4' area and
the `sticky' region just outside corotation around the unstable
`PL5' orbit. Such `stickiness' is observed in many energy levels and
it is responsible for supporting various features of the system (QR3
at $t=20T_{hmct}$). In Fig.18c we see the unstable periodic orbit
(black), starting at the point `PL5' of Fig.18a, superimposed to the
backbone of the system QR3 at $t=20T_{hmct}$ (gray). The `sticky'
behaviour of some orbits around the unstable `PL5' orbit supports
the ring type shape.

\begin{figure}
\centering
\includegraphics[width=8.cm,angle=0]
{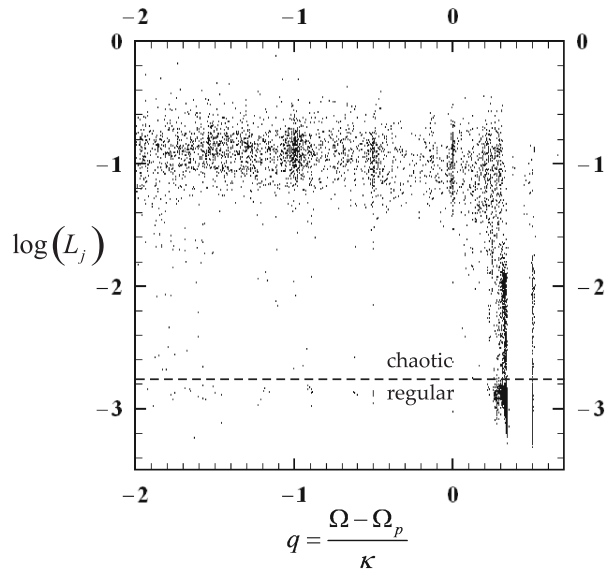} \caption{The logarithm of the Specific Finite Lyapunov
Characteristic Number $(\log L_j)$ for the regular and for the
chaotic orbits as a function of their $q$ value for the experiment
QR3 at $t=20T_{hmct}$.} \label{fig:17}
\end{figure}

Figure 19 shows the evolution of a real chaotic orbit which presents
`stickiness' around the unstable `PL5' orbit of Fig.18a. The
calculated frequency of this orbit is $q\approx 0.0$, i.e. it has
the same angular velocity as the bar and it forms precessing
`bananas'. Such orbits amplify the formation of the ring, which can
be classified as R1, as it is perpendicular to the bar
\citep[see][]{b100}.

\begin{figure*}
\centering
\includegraphics[width=17cm]
{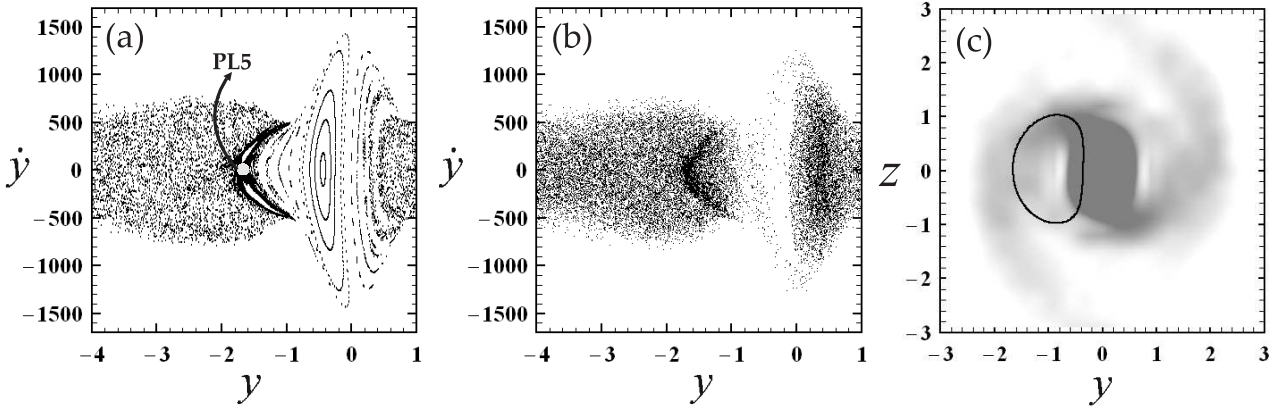} \caption{(a) The 2D surface of section of test particles
for $E_j=-1.1\times10^6$, for the experiment QR3 at t=20$T_{hmct}$.
The position of the unstable periodic orbit PL5 is also plotted (b)
The projection of real n-body particles on the same section as in
(a). (c) The unstable periodic orbit PL5 superimposed on the
backbone of the galaxy at this snapshot (gray background).}
\label{fig:18}
\end{figure*}

In Fig.20 we present the evolution of a real particle in chaotic
orbit, of the experiment QR4 at $t=20T_{hmct}$, in the
configuration space (Fig.20a) as well as in the phase space
($z,\dot{z}$) (Fig.20b). This orbit has $q<-1$ and
$E_j\approx-1.2\times10^6$ (this Jacobi constant lies between
$E_{L_1}$ and $E_{L_4}$). We see that the orbit on the
configuration space is located outside corotation, filling the
area all along the spiral arms (Fig.20a). In Fig.20b the SOS
($z,\dot{z}$) is presented for 50 iterations of test particles,
for $E_j=-1.2\times10^6$, only for the area outside corotation.
The integration time corresponds to 1.5-2.0 Hubble times. The
`stickiness' along the asymptotic curves of the periodic orbits is
evident. In this value of the Jacobi constant the main unstable
periodic orbit is the short period orbit `PL1' or `PL2', around
the Lagrangian points $L_1$ and $L_2$. We note that the density of
points drops abruptly outside the radius that corresponds
approximately to the end of the spiral arms. As there are no
islands of stability or obvious cantori, the only possible reason
of stickiness could be the trapping along the asymptotic curves of
the unstable orbits. In Fig.20b the projections of 350 iterations
of the real 3D chaotic orbit (black dots), are superimposed. We
notice that these sections stay located inside the radius of the
spiral arms. In general, the stickiness of the chaotic orbits
along the asymptotic curves of the unstable curves of the short
period orbit `PL1' or `PL2' as well as all the other curves from
unstable periodic orbits of greater multiplicity, is responsible
for the support of features like rings and spiral arms. In
particular, the invariant manifolds from all the unstable periodic
orbits (for values of $E_j$ between $E_{L_1}$ and $E_{L_4}$)
cannot intersect each other. Thus they are forced to follow nearly
parallel paths in the phase space, and this enhances the
structures supported by them, i.e. the rings and the spiral arms.
A thorough study of the role of asymptotic curves in the
maintenance of spiral structure is done in \citet{b10} and
\citet{b101}. They showed that the apocentres of all the chaotic
orbits with initial conditions along the manifolds corresponding
to the unstable periodic orbits with Jacobi constants $E_{L_1}\leq
E_j\leq E_{L_4}$ create an invariant locus supporting the spiral
arms. In Fig.20c we plot only the apocentres of the orbit plotted
in Fig.20a (black dots in Fig.20c), where we observe that they
coincide very well with the outer parts of the ring and with the
spiral arms (gray background).

\begin{figure*}
\centering
\includegraphics*[width=18cm]
{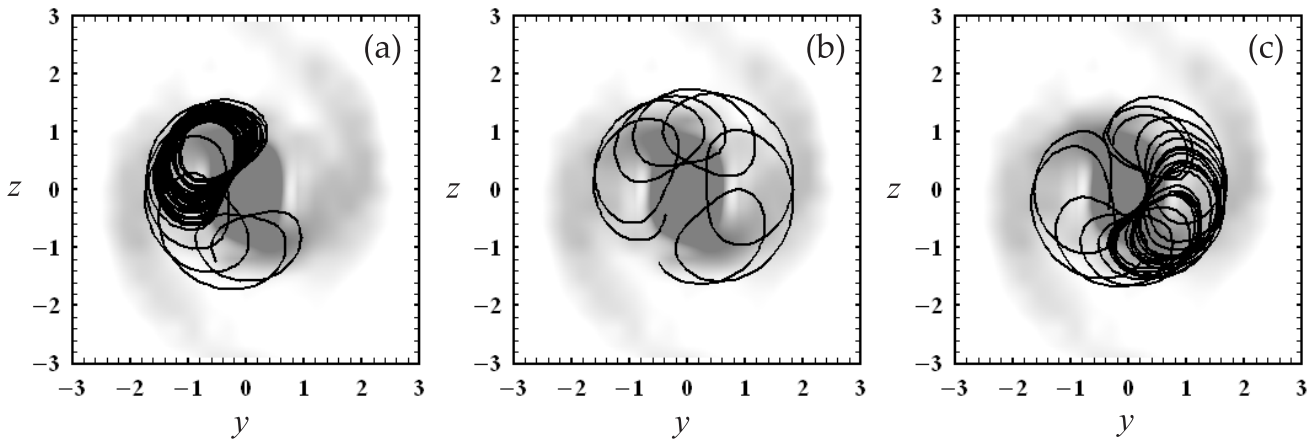} \caption{The evolution of a real chaotic orbit which
presents `stickiness' around the unstable `PL5' of Fig.18a, for the
experiment QR3 at $t=20T_{hmct}$. The calculated frequency of this
orbit is $q\approx 0$, i.e. it has the same angular velocity with
the bar of the system and forms precessing `banana-like' orbits that
amplify the formation of the ring of type `R1'. Compare this figure
with Fig.12f} \label{fig:19}
\end{figure*}

\begin{figure*}
\centering
\includegraphics*[width=17cm,angle=0]
{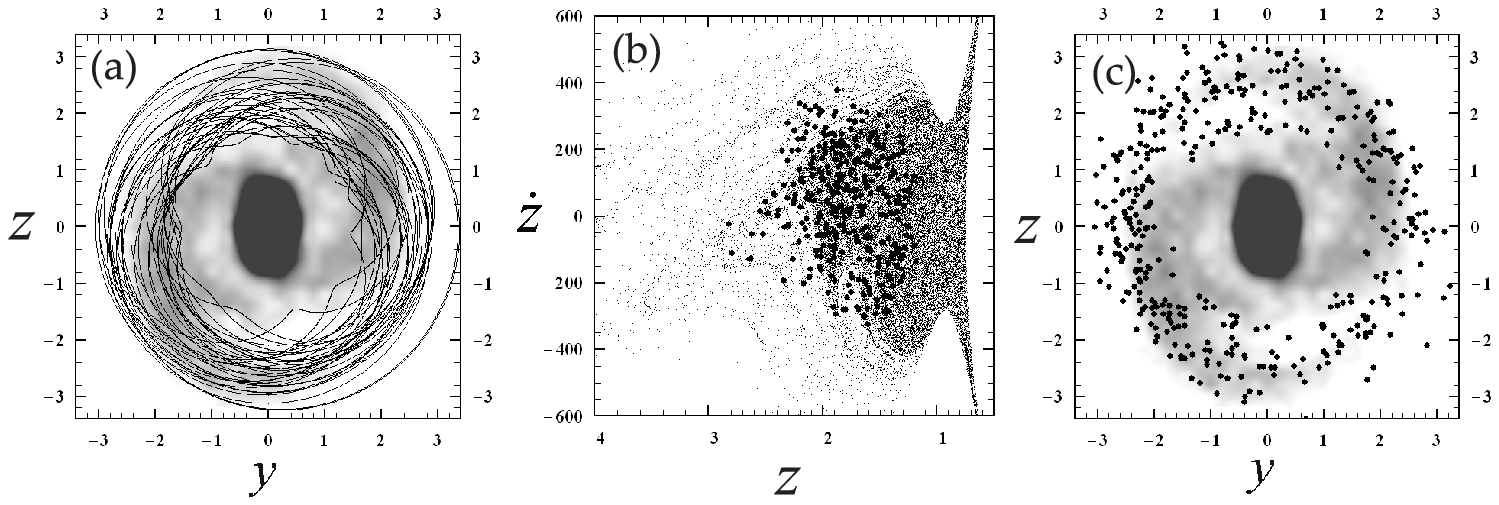} \caption{(a) The evolution of a real chaotic orbit
having $q<-1$ and $E_j\sim-1.2\times10^6$, for the experiment QR4
at $T_{hmct}=20$ (b) The $z \dot{z}$ surface of section for
$E_j=-1.2\times10^6$ for 50 iterations of test particles of the
same experiment. The `stickiness' along the asymptotic curves of
the periodic orbits is obvious. We plot superimposed 350 surfaces
of section corresponding to the real chaotic orbit (black dots)
plotted in (a). (c) Only the apocentres (black dots) of the orbit
in (a) integrated for $\sim$10 Hubble times superimposed on the
backbone of the galaxy (gray background). We notice the good
coincidence with the outer parts of the ring and the spiral arms.}
\label{fig:18}
\end{figure*}

\section{CONCLUSIONS}

We present a detailed investigation of the orbital structure of four
different $N$-body experiments simulating barred-spiral galaxies
exhibiting significant secular evolution within one Hubble time. The
study has been made for two different snapshots during the evolution
of the systems, an early snapshot (t=20$T_{hmct}$) and a late one in
which the systems are close to equilibrium (t=300$T_{hmct}$). The
amplitude of the bar is large in all the experiments and the spin
parameter has a value close to the one of our Galaxy.

The main conclusions of our study are the following:

(i) The ellipticity profile of the bar's isophotes evolves from
flat, at t=20$T_{hmct}$, to declining at t=300$T_{hmct}$, when the
systems are close to equilibrium.

(ii) The boxiness of the bar's isophotes and the pattern speed
($\Omega_p$) of the bar are correlated, i.e the boxiness decreases
with increasing $\Omega_p$.

(iii) Using frequency analysis we have found the main resonances
around which the $N$-body particles in regular orbits are
concentrated. The particles in the chaotic orbits are more
scattered in their frequency maps, however we can still detect
concentrations around a variety of resonances, larger than in the
case of the particles in regular orbits.

(iv) Almost the whole mass component in regular orbits are
particles located inside corotation. These particles support the
bar. The most populated type of resonant regular orbit is the 2:1
or `x1' type of orbit. Another important group of regular orbits
corresponds to the 3:1 resonance at t=20$T_{hmct}$ (except from
the experiment QR1). This gradually transforms to a group `A' type
of orbit at t=300$T_{hmct}$. Group `A' is responsible for the
boxiness of the bar's isophotes.

(v) Group `B' corresponds to retrograde `x4' type of orbits and
exhibits a very small component both in regular and chaotic
orbits, which becomes almost negligible in experiments with great
values of the bar's pattern speed $\Omega_p$.

(vi) Chaotic orbits exist only beyond a threshold of values of the
Jacobi constant value $E_j$. An important percentage of them are
located inside corotation near the resonances 2:1, 3:1, 4:1 and
the group `A' and support the outer regions of the bar. This
component of the chaotic population has the smallest values of the
Specific Finite Time Lyapunov Characteristic Number among all the
chaotic orbits. Therefore, these orbits can be considered as
weakly chaotic, resembling to regular orbits even when they are
integrated for times much longer than a Hubble time.

(vii) Comparing 2D surfaces of sections for test particles with
projections on the same surfaces of sections for real $N$-body
particles we conclude that the areas corresponding to the `x4'
type of orbits are depopulated. The chaotic domain in the surface
of section, inside corotation, are limited by invariant curves in
the 2D approximation of test particles. The same seems to be true
for the projections on the surface of section of the real $N$-body
chaotic orbits, despite the third dimension.

(viii) There are chaotic domains in the surface of section, just
outside corotation, that show `stickiness' due to the existence of
asymptotic manifolds of some unstable periodic orbits. The
stickiness is observed in both the test particles and the real
$N$-body particles. These domains correspond to chaotic orbits that
present a very slow diffusion before escaping from their systems,
and therefore they are responsible for the maintenance of the
general features of the systems such as rings and spiral arms.

(ix) The corresponding frequencies responsible for the formation
of the `R1' rings around the bar, in all our experiments have
values around $q=0$ and $q=-\frac{1}{2}$, while the chaotic orbits
that are responsible for the spiral arms have frequencies
$q\leq-1$.

\section*{Acknowledgments}
We would like to thank professor G. Contopoulos and Dr. C.
Efthymiopoulos for useful discussions and for a careful reading of
the manuscript with many suggestions for improvement. Finally, we
would like to thank the referee for his useful remarks that helped
us to improve our paper.

\label{lastpage}

\end{document}